\documentstyle[12pt]{article}

\begin{document}

\title{GEOMETRIC CONSTRAINT IN BRANE-WORLD}
\author{EDMUNDO M. MONTE\thanks{e-mail: edmundo@fisica.ufpb.br}\\
Departamento de Fisica, Universidade \\Federal da Paraiba, 58059-970, Jo\~{a}o Pessoa, Paraiba, Brazil.}

\maketitle

\begin{abstract}
The brane-worlds model was inspired partly by Kaluza-Klein's theory, where the gravitation and the gauge fields are obtained of a geometry of higher dimension (bulk). Such a model has been showing positive in the sense of we find perspectives and probably deep modifications in the physics, such as: Unification in a scale TeV, quantum gravity in this scale and deviation of Newton's law for small distances. One of the principles of this model is to suppose a space-time embedded in a bulk of high dimension. In this note it is shown, basing on the theorem of Collinson-Szekeres, that the space-time of Schwarzschild cannot be embedded locally and isometrically in a bulk of five dimensions with constant curvature,(for example ADS-5). From the point of view of the semi-Riemannian geometry this last result consists constraints to the model brane-world.
\end{abstract}

\section{Introduction}

Recent theories involve the idea that the 3-spatial dimensions in which we live could be a 3-spatial-dimensional `membrane' embedded in a much larger extra dimensional space, and that the hierarchy is generated by the geometry of the additional dimensions. Such ideas have led to extra dimensional theories which have verifiable consequences at the TeV scale \cite{Hewett}.

Randall and Sundrum have proposed an interesting scenario of extra non-compact dimensions in which four-dimensional gravity emerges as a low energy effective theory, to solve the hierarchy problems of the fundamental interactions. This proposal is based on the assumption that ordinary matter and its gauge interactions are confined within a four dimensional hypersurface, the physical brane, embedded in a five-dimensional space and the fact that a bound state of a five-dimensional graviton exists and is localized near the physical brane\cite{RS2}.

The emphasis in the development of higher dimensional theories has recently shifted toward  the brane-world picture\cite{ADD}. A brane-world may be regarded as a space-time locally embedded in a higher dimensional space, the bulk,  solution of higher dimensional Einstein's equations. Furthermore, the embedded geometry  is assumed to  exhibit quantum  fluctuations with respect to  the extra dimensions at the  TeV scale of  energies. Finally, all  gauge interactions belonging to the standard model must remain confined to the four-dimensional space-time. Contrasting with  other  higher dimensional theories, the extra dimensions may be large and even infinite, with the possibility of being observed by  TeV accelerators. The embedding  conditions relate the  bulk  geometry  to the brane-world geometry, as it is clear from the  Gauss-Codazzi-Ricci  equations\cite{ME}.

The Randall-Sundrum model consists of a three-brane embedded in a five-dimensional space which is asymptotically anti-de Sitter. A generic form of the metric in this space is
\[
ds^2=e^{-2\kappa|y|}{\bar g}_{\mu\nu}dx^\mu dx^\nu+dy^2,
\]
the four-dimensional metric ${\bar g}_{\mu\nu}$ being asymptotically flat.
The asymptotic metric with ${\bar g}_{\mu\nu}
=\eta_{\mu\nu}$ satisfies the vacuum
Einstein equations
\[
R_{mn}- \frac{1}{2}Rg_{mn}-\Lambda g_{mn}=0
\]
with the cosmological constant $\Lambda=-6\kappa^{2}$, where $\kappa$ describes the curvature scale, which is assumed to be of order the Planck scale. Here we adopt the convention\cite{Giannakis-Ren} that the Greek indices take values $1-4$, the Latin indices $1-5$. The anti-de-Sitter five dimensional space is the space of constant curvature $K_{o}$ that depends of constants $\Lambda$ and $\kappa$. For our objective, without generality loss, we will place that constant as $K_{o}$.

In order to describe the real world, the Randall-Sundrum scenario has to satisfy all the existing tests of General Relativity, with base in the motion of material particles within the Schwarzschild brane that is given by four-dimensional geodesic equation\cite {Duck}.

The purpose of this paper is to show that we can not release the embedding of Schwarzschild brane in the five dimensional bulk, with constant curvature, so constituting a constraint in one of the hypotheses of Randall-Sundrum model, i.e., that a three-brane is embedded in a five-dimensional space which is asymptotically anti-de Sitter.

For fundamental theorem of submanifolds the four dimensional space-time of Schwarzschild with usual metric $g_{\mu\nu}$ can be embedded locally and isometrically in a five dimensional bulk of constant curvature $K_{o}$ if and only if there exists a symmetric tensor  $b_{\mu\nu}$ satisfying respectively the Gauss and Codazzi equations:

\begin{equation}
R_{\mu\nu\alpha\beta}=(b_{\mu\alpha}b_{\nu\beta}-b_{\mu\beta}b_{\nu\alpha})+K_{o}(g_{\mu\alpha}g_{\nu\beta}-g_{\mu\beta}g_{\nu\alpha})
\end{equation}

\begin{equation}
b_{\mu\nu;\alpha}-b_{\mu\alpha;\nu}=0,
\end{equation}

where the (;) is the covariant derivative in relation to $g_{\mu\nu}$ metric, $R_{\mu\nu\alpha\beta}$ is the curvature tensor of the space-time of Schwarzschild and $b_{\mu\nu}$ is the extrinsic curvature this space.

Contracting equation $(1)$ yields, on putting $R_{\nu\alpha}=0$,
\[
(b_{\mu\alpha}b_{\nu}^{\mu}-b_{\mu}^{\mu})=3K_{o}g_{\nu\alpha}.
\]
This equation has been investigated by Szekeres and Collinson in another context \cite{Collinson}. The tensor  $b_{\nu\alpha}$ must take one of the two canonical forms,
\[
(i) b_{\nu\alpha}=\lambda g_{\nu\alpha} \hspace{1cm} -K_{o}=\lambda^{2}
\]
\[
(ii) b_{\nu\alpha}=\lambda g_{\nu\alpha}+2\lambda u_{\nu}u_{\alpha}-2\lambda s_{\nu}s_{\alpha} \hspace{1cm} 3K_{o}=-\lambda^{2},
\]
where $u_{\nu}$ and $s_{\alpha}$ are vectors satisfying
$-u_{\nu}u^{\nu}=s_{\nu}s^{\nu}=1$ \hspace{0,2cm}and \hspace{0,2cm} $u_{\nu}s^{\nu}=0$.
Substituting $(i)$ into equation $(1)$ results $R_{\mu\nu\alpha\beta}=0$,  which contradicts the hypothesis that the Schwarzschild brane is not flat.
The expression in $(ii)$ can be written as\cite{Collinson}:

\begin{equation}
b_{\nu\alpha}=\sqrt{\left|3K_{o}\right|}g_{\nu\alpha}
-\sqrt{\left|12K_{o}\right|}(l_{\nu}n_{\alpha}+n_{\alpha}l_{\nu})
\end{equation}

where the null vectors $l_{\nu}$ and $n_{\nu}$ are defined by
\[
l_{\nu}=(s_{\nu}+u_{\nu})/\sqrt{2}  \hspace{1,5cm} n_{\nu}=(s_{\nu}-u_{\nu})/\sqrt{2}.
\]
Substituting $(3)$ into $(1)$, $(2)$ and after of long calculus obtain the expression
\[
-4K_{o}l_{[\beta}g_{\mu][\alpha}l_{\nu]}=0.
\]
Note that this expression also lead us to absurd, because $K_{o}\neq 0$.

We verify with this result that we don't obtain the embedding of Schwarzschild brane into five dimensional bulk with constant curvature, for example, anti-de Sitter space. So, from the point of view of embedding, in the models of the brane-worlds  exist constraints. Another result proved by Szekeres \cite{Collinson} proves that  we don't obtain the embedding of Schwarzschild brane into five dimensional flat bulk. From other side we can obtain the embedding of Schwarzschild brane into six dimensional bulk flat with different signatures, when there is change in the topology of Schwarzschild brane \cite{SchwTop}.

I would like to thank Professor Marcos Maia, for useful discussions.


\begin{thebibliography}{8}

\bibitem{Collinson}
P. Szekeres, {\it Nuovo Cimento}, {\bf 43}, 1062 (1966); C.D. Collinson, {\it J. Phys.} {\bf A4}, 206 (1971).
\bibitem{Hewett}
J.L. Hewett, {\it Phys. Rev. Lett.} {\bf 82}, 4765 (1999).
\bibitem{RS2}
L. Randall and R. Sundrum,  {\it Phys. Rev. Lett.} {\bf 83}, 3370 (1999).
\bibitem{ADD}
K. Maeda {\it et al.}, {\it Phys. Rev.} {\bf D62}, 24012 (2000).
\bibitem{Giannakis-Ren}
I. Giannakis and H. Ren, {\bf hep-th 0007053} (2000).
\bibitem{Duck}
W. Muck, K. S. Viswanathan and I. V. Volovich, {\bf hep-th/0002132}, (2000); J. Garriga and T. Tanaka, {\it Phys. Rev. Lett.} {\bf 84}, 2778 (2000); M. J. Duf and J. T. Liu, {\it Phys. Lett.} {\bf 476B}, 363 (2000).
\bibitem{ME}
M. D. Maia and E. M. Monte, {\it Phys. Lett.} {\bf A297}, 9 (2002).
\bibitem{SchwTop}
E. M. Monte and M. D. Maia, {\it Int. J. Mod. Phys. A}, {\bf 17}, 4355 (2002).

\end{thebibliography}
\end{document}